\documentclass[conference,10pt]{IEEEtran}
\author{
	\IEEEauthorblockN{Juhwan Yoo\IEEEauthorrefmark{1}, Yao Xie\IEEEauthorrefmark{1}, Andrew Harms\IEEEauthorrefmark{2}, Waheed U. Bajwa\IEEEauthorrefmark{3}, and Robert Calderbank\IEEEauthorrefmark{1}}
	\IEEEauthorblockA{\IEEEauthorrefmark{1}{\small Department of Electrical and Computer Engineering, Duke University, Durham, NC, USA}}
	\IEEEauthorblockA{\IEEEauthorrefmark{2}{\small Deparment of Electrical Engineering, Princeton University, Princeton, NJ, USA}}
	\IEEEauthorblockA{\IEEEauthorrefmark{3}{\small Department of Electrical and Computer Engineering, Rutgers University, Piscataway, NJ, USA\vspace{-0.7cm}}}
}
\usepackage[utf8x]{inputenc}
\usepackage{graphicx,amsmath,amsfonts,amssymb,amsthm,
    cite,url, algorithm, algpseudocode, ifthen}
\usepackage[usenames,dvipsnames]{color}
\usepackage{array}
\usepackage[pdftex,pdfborder={0 0 0}]{hyperref} 
\usepackage{fullpage}
\usepackage{verbatim}
\usepackage[font=footnotesize,caption=true]{caption,subfig}
\usepackage{wrapfig}
\usepackage{epsfig}
\usepackage{psfrag}
\usepackage[normalem]{ulem}     
\usepackage{cancel} 
\usepackage{MnSymbol}
\newlength{\myCaptionSize}
\newlength{\myFigSize}
\newlength{\mySmallFigSize}
\DeclareMathAlphabet{\mathbbb}{U}{bbold}{m}{n}

\newcommand{\C}{\mathbb{C}} 		

\newcommand{\norm}[1]{{\left\lVert{#1}\right\rVert}}

\newcommand{\e}{\mathrm{e}}

\newcommand{\x}{\ensuremath{\mathbf{x}}}
\newcommand{\y}{\ensuremath{\mathbf{y}}}

\newcommand{\beq}{\begin{equation}}
\newcommand{\eeq}{\end{equation}}
\newcommand{\bea}{\begin{eqnarray}}
\newcommand{\eea}{\end{eqnarray}}

\newcommand{\beqsplit}{\begin{equation}\begin{split}}
\newcommand{\eeqsplit}{\end{split}\end{equation}}

\newcommand{\mb}[1]{\ensuremath{\mathbf{#1}}}

\newtheorem{thm}{Theorem}
\newtheorem*{thm*}{Theorem}

\newtheorem*{lem*}{Lemma}
\newtheorem{definition}{Definition}
\newtheorem*{definition*}{Definition}

\newtheorem*{prop*}{Proposition}

\newtheorem*{notation*}{Notation}

\newtheorem*{cor*}{Corollary}

\newtheorem*{remark*}{Remark}

\newtheorem{Q*}{Question}

\newtheorem*{term*}{Terminology}

\newtheorem*{key*}{Keywords}

\newtheorem*{mynote*}{{\textcolor{red}{note:}} }

\newtheorem*{question*}{{\textcolor{blue}{question:}} }

\newtheorem*{answer*}{{\textcolor{green}{answer:}} }

\newtheorem*{todo*}{{\textcolor{Magenta}{To Do:}} }

\newtheorem*{claim*}{claim:}

\newtheorem*{ass*}{assumption:}

\newtheorem*{principle*}{Principle}

\numberwithin{equation}{section}

\newcommand{\bitem}{\begin{itemize}}
\newcommand{\eitem}{\end{itemize}}
\newcommand{\benum}{\begin{enumerate}}
\newcommand{\eenum}{\end{enumerate}}
\newcommand{\bdesc}{\begin{description}}
\newcommand{\edesc}{\end{description}}

\newcommand{\bverb}{\begin{verbatim}}
\newcommand{\everb}{\end{verbatim}}

\newcommand{\iii}{\ensuremath{\mathit{i}}}

\newcommand{\togglesection}[1]{\ifthenelse{\boolean{#1}}}
\newcommand{\btoggle}[1]{\ifthenelse{\boolean{#1}}}

\newboolean{printresearch}
\setboolean{printresearch}{true}

\newboolean{printpersonal}
\setboolean{printpersonal}{true}

\newboolean{printcomments}
\setboolean{printcomments}{true}

\newcommand{\ben}{\begin{eqnarray}}

\newcommand{\een}{\end{eqnarray}}

\newcommand{\vect}{\boldsymbol} 
\newcommand{\SNR}{\textsf{SNR}}

\newcommand{\ab}{\vect{a}}

\newcommand{\xb}{\vect{x}}

\newcommand{\zb}{\vect{z}}
\newcommand{\A}{\vect{A}}

\newcommand{\X}{\vect{X}}
\newcommand{\I}{\vect{I}}

\newcommand{\yb}{\vect{y}}

\newcommand{\wb}{\vect{w}}

\newcommand{\FDP}{\textsf{FDP}}

\newcommand{\LAR}{\textsf{LAR}}

\newlength{\maxcolumnfigwidth}
\divide\maxcolumnfigwidth by 2

\newlength{\maxfigwidth}
\title{Finding Zeros: Greedy Detection of Holes}
\begin{document}
\maketitle
\thispagestyle{plain}
\pagestyle{plain}
\begin{abstract}
In this paper, motivated by the setting of white-space detection~\cite{pcast}, we present theoretical and empirical results for detection of the zero-support $\mathcal{E}$ of $\xb \in \mathbb{C}^p$ ($x_i = 0$ for $i \in \mathcal{E}$) with reduced-dimension linear measurements. We propose two low-complexity algorithms based on one-step thresholding~\cite{bajwaGabor2010} for this purpose. The second algorithm is a variant of the first that further assumes the presence of group-structure in the target signal~\cite{bajwaGroup} $\xb$. Performance guarantees for both algorithms based on the worst-case and average coherence (group coherence) of the measurement matrix is presented along with the empirical performance of the algorithms.
\end{abstract}
%
\begin{keywords}
Zero-Detection, White-Space Detection, Compressed-Sensing, Dimensionality-Reduction, Average Coherence, Average Group Coherence.
\end{keywords}
\section{Introduction}\label{sec:intro}
The principal idea that underlies research in the area of big data~\cite{bigdataScience} is that the majority of information in many signals of interest is structured and therefore lies in a much lower dimensional subset of the ambient signal dimension.  This idea was first popularized by the theory of Compressed-Sensing~\cite{CRT2006} (CS) which demonstrated that a vector $\x \in \mathbb{R}^p$ with $k$-sparse  ``non-zero'' support ($\norm{\x}_0 =k$) could be recovered with $n = \mathcal{O}(k \log(p))\ll p$ non-adaptive linear measurements $\y = \A\x$,   
 where $\A\in \C^{n \times p}$. The initial results prescribed the use of random sensing matrices and signal recovery via solving an LP which finds, among all solutions consistent with the measurements, the one with minimum $\ell_1$ norm. The advent of CS inspired a large amount of research in areas related to dimensionality reduction (DR) with goals spanning: exploiting different kinds of structure~\cite{bajwaGroupThresholding}, reduced-dimension signal processing~\cite{yaoMUDjournal2013}, structured sensing matrix design~\cite{rauhutSCS,sirjsp2012}, and efforts at employing its results~\cite{rmpi_rfic}.

While much work has been done within the DR framework, one area that has remained relatively unexplored is the detection of zeros: given measurements found in the standard CS setup, 
we are interested in detecting the support of the entries of $\x$ that are equal to zero. Philosophically, the goal of finding zeros    
 can be interpreted as detecting absence/non-existence. This goal can be found in many resource-allocation applications where the goal is to cheaply query a system of interest to determine what is not being used or not working.
 One conspicuous example where this goal manifests itself is white-space detection~\cite{harmsRapid2012}. White-space detection is a sub-problem of the efficient spectrum sensing problem whose goal is to more efficiently use large swaths of bandwidth by designing spectrum sensors that quickly find and opportunistically communicate over $\emph{unused}$ pieces of spectrum. Many research efforts with the aim of addressing this problem have been heavily influenced by CS-like ideas in recent years. A common strategy is to use the sparse-approximation/random sampling machinery and recover the entire spectrum (or its support) to determine the location of unoccupied channels. 
 This approach, given the goal of finding free channels to transmit across, is inefficient in several respects. The first is that it solves an estimation problem to what is intrinsically a detection problem. While exact knowledge of spectrum usage is ideal, it is often sufficient and less costly to obtain a large subset of the locations of unused pieces of spectrum. 
In particular, more efficient detection of unused pieces of spectrum can become critically important in situations where the system is required to quickly adapt, e.g., the support is changing rapidly.  The second is that spectrum usage exhibits group behavior, i.e., use of one portion of spectrum is often indicative of activity in other portions of the spectrum.  For example, the entirety of spectrum is broken up into channels and most of a channels bandwidth will likely be active at once.
	
In this paper, drawing inspiration from the setting of white-space detection, we are concerned with a specific type of zero detection problem: detection of a large \emph{subset} of non-zero elements, without requiring complete or exact support/zero pattern recovery. An additional goal is to design algorithms that have low complexity and that are amenable to use in an adaptive setting.
In this spirit, we present two algorithms in Sec.~\ref{sec:algorithms} that utilize methods and results from work in support detection~\cite{bajwaGabor2010, harmsRapid2012} and group model selection~\cite{bajwaGroupThresholding}. The first algorithm (Alg.~\ref{alg:OMPbased}) is a simple modification of one-step thresholding  (OST)~\cite{bajwaGabor2010} 
and the second (Alg.~\ref{alg:GroupOMPbased}) is an extension of OST in the setting of group model selection in~\cite{bajwaGroup}. The performance guarantees for these algorithms are presented in Sec.~\ref{sec:guarantee}. The proof of the guarantees is given in Appendices~\ref{sec:proof1} and \ref{sec:proof2}. The proofs utilize the concepts of: average coherence/group coherence ($\nu,\nu^g$), worst-case coherence/group coherence ($\mu,\mu^g$), the statistical orthogonality condition (StOC), and the coherence property (CP)~\cite{bajwaGabor2010,bajwaAvgCoherence}. Numerical simulations of the two algorithms are presented in Sec.~\ref{sec:numerical} and the paper concludes in Sec.~\ref{sec:conclusion}. 
\section{Problem Formulation and Algorithms}\label{sec:formulation}
Let $\xb \in \C^p$ where $\norm{\xb}_0 = k$. Denote the zero-support of $\xb$ with $\mathcal{E} \subset \{1, \dots,p\}$ and its complement $\mathcal{I}=\mathcal{E}^c$: $x_i = 0$ for $i \in \mathcal{E}$. The two zero-detection algorithms presented in this paper generate estimates of the zero-support ($\hat{\mathcal{E}}$) for
the following two measurement models corresponding to the presence/absence of group-structure in $\xb$.
\subsection{Measurement Models}
\subsubsection{Non-group-structure model}
The non-group-structure setting corresponds to the standard model of CS given by
\ben\label{model1}
\yb = \A\xb + \wb, 
\een
where $\yb \in \mathbb{C}^{n\times 1}$ is the measurement vector, $\A\in \mathbb{C}^{n\times p}$ ($n \ll p$) is the measurement matrix with unit-norm columns, $\xb \in \C^{p}$ is the signal vector ($k = \norm{\xb}_0$), and  $\wb\sim\mathcal{N}(0, I \sigma^2)$.
  The zero-detection algorithm corresponding to this setting is Alg.~\ref{alg:OMPbased}.
\subsubsection{Group-structure model}
The group structure model corresponds to scenarios, such as statistical model selection, where the existence of a single entry in $\xb$ implies the presence of other related entries in the true model. In this paper we examine situations where there are $q$ groups with each group consisting of $r$ entries of $\xb$. In this case, we modify model (\ref{model1}) to
\ben
\yb = \sum_{i=1}^q \A_i \xb_i + \wb = 
\sum_{i \in \mathcal{K}} \A_i \xb_i + \wb,
\een
where $\A_i \in \mathbb{C}^{n\times r}$ is a sub matrix of $\A$, and $\xb_i$ are the coefficients associated with group $i$. Let the set $\mathcal{K}\triangleq \{1\leq i \leq q: \xb_i \neq \vect{0}\}$ denote the true underlying model with $k \triangleq |\mathcal{K}|$ groups that have non-zero coefficients. When discussing group-structure, $\mathcal{E}$ will denote the indices of \emph{groups} that have zero coefficients. The zero-detection algorithm corresponding to this setting is Alg.~\ref{alg:GroupOMPbased}.
\subsection{Zero detection algorithms}\label{sec:algorithms}
Both zero-detection algorithms~\ref{alg:OMPbased} and \ref{alg:GroupOMPbased} generate an estimate of zero-support of $\xb$ ($\hat{\mathcal{E}}$) by applying the Hermitian transpose of the measurement matrix $\A^H$ to the output measurements and retaining the indices of the $\theta = |\hat{\mathcal{E}}|$ lowest magnitude coefficients. Intuitively, the underlying idea behind this operation is similar to that of orthogonal matching pursuit (OMP)~\cite{troppOMP} and OST in that 
it expresses the belief that low correlation of the output with the $i^{\text{th}}$ column of the measurement matrix ($\mb{a}_i$) is indicative of the fact that $x_i = 0$.
\begin{algorithm}
\caption{Zero-Detection One-Step Thresholding (ZD-OST)}
\begin{algorithmic}[1]
\State Input: measurements $\yb$, design matrix $\A$, number of empty bands to select $\theta$
 \State Initialization: $\hat{\mathcal{E}}=\{\emptyset\}$
\State Obtain measurements and apply processing matrix: $\mb{s}=\X^H \yb$.
 \State Sort $|s_i|$ in ascending magnitude and assign to $\hat{\mb{s}}$.
 \State Construct set of lowest $\theta$ magnitudes $\hat{\mathcal{E}} = \hat{\mb{s}}(1:\theta)$.
 \State Output $\hat{\mathcal{E}}$.
\end{algorithmic}
\label{alg:OMPbased}
\end{algorithm}\vspace{-0.2cm}
\begin{algorithm}
\caption{Zero-Detection Group Thresholding (ZD-GroTh)}
\begin{algorithmic}[1]
\State Input: measurements $\yb$, design matrix $\A$, size of the group $r$, number of empty groups to select $\theta$
 \State Initialization: $\hat{\mathcal{E}}=\{\emptyset\}$
\State Obtain measurements and apply the following: $s_i=\|\X_i^H \yb\|_2$.
 \State Sort $|s_i|$ in ascending magnitude and assign to $\hat{\mb{s}}$.
 \State Construct set of lowest $\theta$ magnitudes $\hat{\mathcal{E}} = \hat{\mb{s}}(1:\theta)$.
 \State Output $\hat{\mathcal{E}}$.
\end{algorithmic}
\label{alg:GroupOMPbased}
\end{algorithm}\vspace{-0.2cm}
\section{Performance Guarantees}\label{sec:guarantee}
Since we are interested in estimating sets $\hat{\mathcal{E}}$ that with high probability contain subsets of the zero-support, the metrics with which we establish performance guarantees for 
ZD-OST and ZD-GroTh are the false-discovery proportion (FDP) 
\ben\label{eq:fdp}
\textsf{FDP}(\hat{\mathcal{E}}) \triangleq \frac{|\hat{\mathcal{E}}\backslash\mathcal{E}|}{|\hat{\mathcal{E}}|},
\een
as well as the probability of error $\text{P}_{\text{e}}=\mathbb{P}\{\hat{\mathcal{E}}\cap \mathcal{E}=\emptyset\}$.
\subsection{Performance Guarantees for ZD-OST}
In order to establish performance guarantees for ZD-OST,
it is necessary to define several quantities and review a few concepts central to the main arguments. 
Let $x_{(m)}$ denote the $m^{\text{th}}$ largest magnitude non-zero entry of $\xb$. Hence $|x_{(1)}| \geq |x_{(2)}| \geq \ldots \geq |x_{(k)}|$.
Define the signal-to-noise ratio (SNR), the $m^{\text{th}}$ largest-to-average ratio ($LAR_{m}$), and the minimum SNR ($\text{SNR}_{\text{min}}$) as 
\begin{equation}
\begin{split}
&\SNR \triangleq \frac{\|\xb\|_2^2}{\mathbb{E}[\|\wb\|_2^2]}, \quad\LAR_m \triangleq \frac{\|x_{(m)}\|^2}{\|\xb\|_2^2/k},\quad m = 1, \ldots, k,\\
&\SNR_{\min} \triangleq \frac{x_{\min}^2}{\sigma^2},
\end{split}
\end{equation}
respectively.
In addition, we define two coherence properties of the unit-column norm matrix $\A$: the worst-case coherence (Eq.~\ref{eq:wccoherence}) and the average coherence (Eq.~\ref{eq:avgcoherence}) 
\begin{align}
\mu(\A) &\triangleq \max_{i\neq j} |\ab_i^H \ab_j|, 
\label{eq:wccoherence}\\
\nu(\A) & \triangleq \frac{1}{p-1} \max_i \left|
\sum_{j:j\neq i} \ab_i^H \ab_j \right|, 
\label{eq:avgcoherence}
\end{align}
We further define the statistical orthogonality condition (StOC).
\begin{definition}[Statistical Orthogonality Condition~(Def. 3~\cite{bajwaGabor2010})] 
Let $\bar{\Pi} = (\pi_1, \ldots, \pi_p)$ be a random permutation of $\{1,\ldots, p\}$, and define $\Pi \triangleq (\pi_1, \cdots, \pi_k)$, and $\Pi^c \triangleq (\pi_{k+1}, \ldots, \pi_p)$ for any $k \in \{1, \ldots, p\}$. Then the $n\times p$ normalized design matrix $\A$ is said to satisfy the ($k,\epsilon,\delta$)-statistical orthogonality condition (StOC) if there exist $\epsilon,\delta \in [0, 1)$ such that the inequalities 
\begin{align}\label{eq:stoc}
\|(\A_\Pi^H \A_\Pi - \I)\zb\|_\infty &\leq \epsilon \|\zb\|_2 \\
\|\A_{\Pi^c}^H \A_\Pi \zb\|_\infty & \leq \epsilon \|\zb\|_2
\end{align}
hold for every fixed $\zb \in \mathbb{C}^k$ with probability exceeding $1-\delta$ with respect to the random permutation $\bar{\Pi}$. 
\end{definition}
Having established the above conventions, we can now present the following theorem for the performance of ZD-OST.
\begin{thm}
Assume that the noise is $\wb \sim \mathcal{CN}(0, \sigma^2)$, and $\mu = \frac{\mu_0}{\sqrt{\log p}}$ for some constant $\mu_0 > 0$. Also assume that $\SNR_{\min} > 16 \log p$.
\begin{enumerate}
\item   Let $\epsilon_0 = (\sqrt{\SNR_{\min}}-4\sqrt{\log p})/(2\sqrt{\SNR}) > 0$. When $\theta = 1$, if 
\ben
k<\min\left\{\left(\frac{\epsilon_0-4(2+a^{-1})\mu_0}{\nu}\right)^2, (1+a)^{-1}p\right\}, \label{k_bound}
\een
for some $a > 1$, 
then  
$\text{P}_{\text{e}}\leq \sqrt{2/\pi}p^{-1} + 4p^{1-\alpha}$, where $\alpha  = (\epsilon_0-\sqrt{k}\nu)^2/(c\mu_0^2) > 1$.
\item If (\ref{k_bound}) holds, then we have that with probability exceeding $1-\sqrt{2/\pi}p^{-1} - 4p^{1-\alpha}$
\begin{equation}
\begin{split}
\FDP(\hat{\mathcal{E}}) & \leq (k-m) /\theta,
\end{split}
\end{equation}
where $m$ is the largest integer for which the following is true:
\[\LAR_{(m)} \geq 
\max\left\{
\frac{c_1k\log p}{n\SNR}, c_2 \mu^2\log p \right\},\]
with $c_1 = 32t^{-1}$, $c_2 = 800(1-t)^{-1}$ for some $t\in (0, 1)$.
\end{enumerate}
\end{thm}
{\em Remarks:} To interpret the results in Theorem 1, (\ref{k_bound}), since $\SNR_{\min} > 16 \log p$, we can choose $x_{\min}^2  = (1+\gamma) 16\sigma^2 \log p$ for some constant $\gamma > 1$. For this choice, $\SNR = (k/n)(1+\gamma)16 \log p$, and $\epsilon_0 = \sqrt{n/k}(\sqrt{1+\gamma}-1)/(2\sqrt{1+\gamma})$. In the high SNR regime, $\gamma\rightarrow \infty$, $\epsilon_0 \rightarrow (1/2)\sqrt{n/k}$, and hence the first term in (\ref{k_bound}) tends to $((1/2)\sqrt{n/k} - 4(2+a^{-1})\mu_0)^2/\nu^2$, and when $n/k > 64(2+a^{-1})\mu_0^2$, this is approximately $n/(4k\nu^2)$. This demonstrates that if $\nu$ is sufficiently small, the first term in (\ref{k_bound}) is not binding, which implies that we may not need $k$ to be very small relative to $n$. 
This is also demonstrated by the numerical experiments in Sec.~\ref{sec:numerical} that show successful recovery of large subsets of zero even in the absence of sparsity. 
\subsection{Performance Guarantee for ZD-GroTh}
In order to present performance guarantees for group thresholding, we will need to introduce a few additional concepts. First, we define the group-structure analogues of Eqs.~\ref{eq:wccoherence} and~\ref{eq:avgcoherence}: the worst-case group coherence and the average group coherence as
\ben\label{eq:wcgcoherence}
&\mu^g \triangleq \max_{i\neq j, i, j \in \{1, \ldots, q\}}
\|\A_i^H \A_j\|_2, \\\quad
&\nu^g \triangleq \frac{1}{q-1} \max_{i=1, \ldots, q}
\|\sum_{j=1, \ldots, q, j\neq i} \A_i^H \A_j\|_2.
\een
In addition, we define the group coherence property
\begin{definition}[The Group Coherence Property~(Def. 1~\cite{bajwaGroup})]
 The $n\times rq$ measurement matrix $\A$ is said to satisfy the group coherence property if the following two conditions hold for some positive constants $c_{\mu}$ and $c_{\nu}$:
\ben\label{eq:grpcoherenceproperty}
\mu^g \leq \frac{c_\mu}{\sqrt{\log q}}, \quad
\nu^g \leq c_\nu \mu^g \sqrt{\frac{r\log q}{n}}.
\een 
\end{definition}
Let $\xb_{(i)}$ to be the $i^{\text{th}}$ largest group of non-zero coefficients: $\|\xb_{(1)}\|_2 \geq \|\xb_{(2)}\|_2 \geq \ldots \geq \|\xb_{(k)}\|_2 > 0$. 
The following theorem is adapted from (Theorem 1, \cite{BajwaMixon2013}):
\begin{thm}
Suppose $\A$ satisfies the group coherence property with parameters $c_\mu$ and $c_\nu$. Fix parameter $c_1 \geq 2$, $c_2 \in (0, 1)$, and define parameters $c_3 \triangleq [32\sqrt{2e}(2c_1 -1)]/[(1-c_2)(c_1-1)]$. Then, under the assumptions $c_1 r k\leq n$, $c_\mu < c_3^{-1}$, and $c_\nu \leq \sqrt{c_1}c_2 c_3$, we have that with probability at least $1-(1+e^2) q^{-1} $ that 
$\FDP(\widehat{\mathcal{K}})\leq (k-m)/\theta$, where $m$ is the largest integer for which the inequality $\|\xb_{(m)}\|_2 \geq c_3 \mu^g \|\xb\|_2 \sqrt{\log q} + 2\sigma\sqrt{2\log q + r/2 \log2}$ holds. The probability is with respect to the uniform distribution of the true model $\mathcal{K}$ over all possible models. 
\end{thm}
\section{Numerical Experiments}\label{sec:numerical}
\begin{figure}[h!]
\centering
\includegraphics[width=0.26\textwidth]{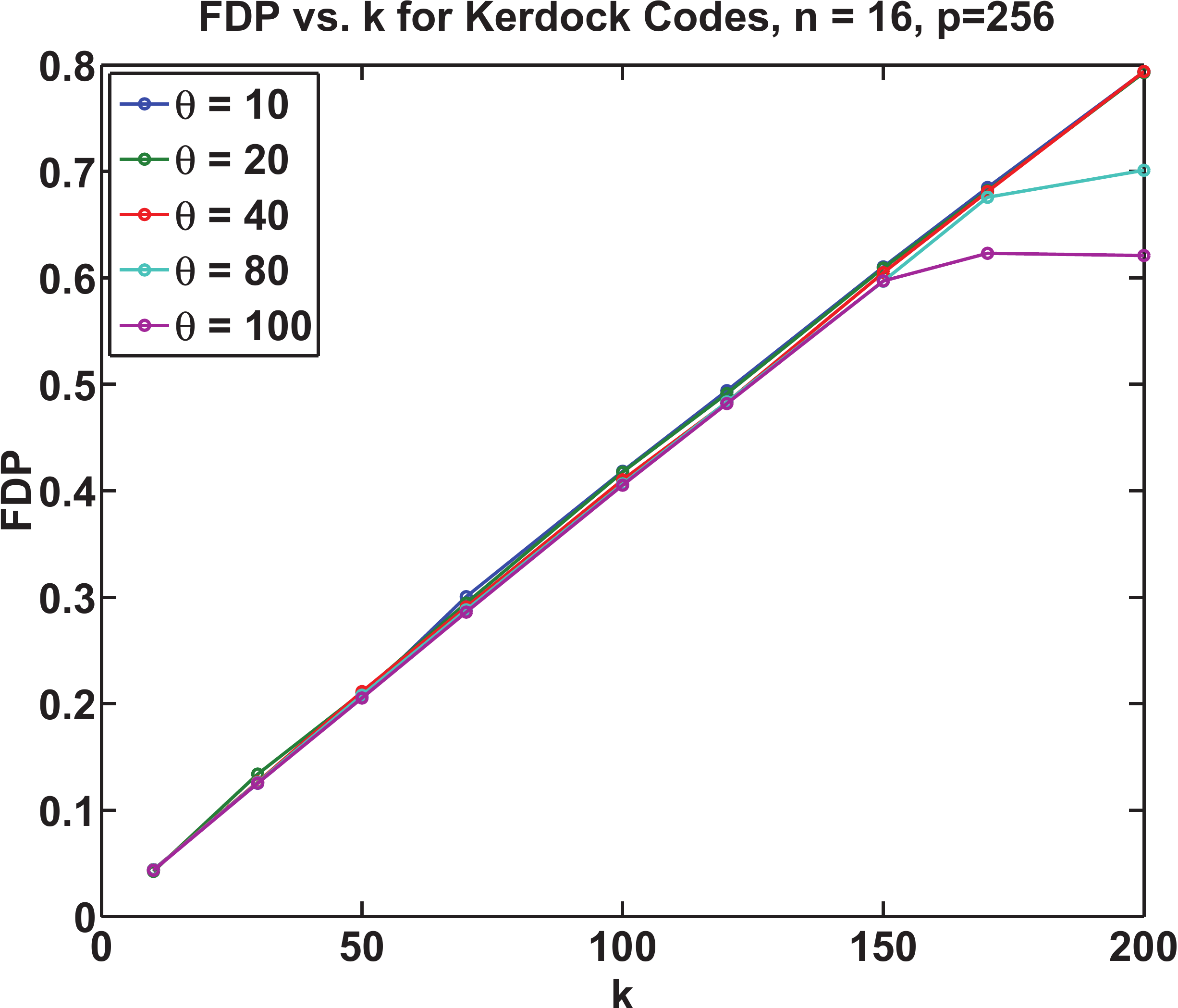}
\caption{The FDP of Kerdock codes as a function of $k$ for several values of $\theta$. Each data point represents the average of 5000 trials. The amplitudes $|x_{\ell}|$ were uniformly distributed in $[1, 1000]$ and $\sigma^2 = 500$. Note, in cases where $\theta > |\mathcal{E}|=p-k$ we plotted the quantity $|\hat{\mathcal{E}}\cap \mathcal{E}|/|\mathcal{E}|$ to represent the total fraction of the zero-support recovered \label{fig:nongroupkerdockFDP}.}
\end{figure}

\begin{figure}[h!]
\begin{center}
\includegraphics[width = 0.26\textwidth]{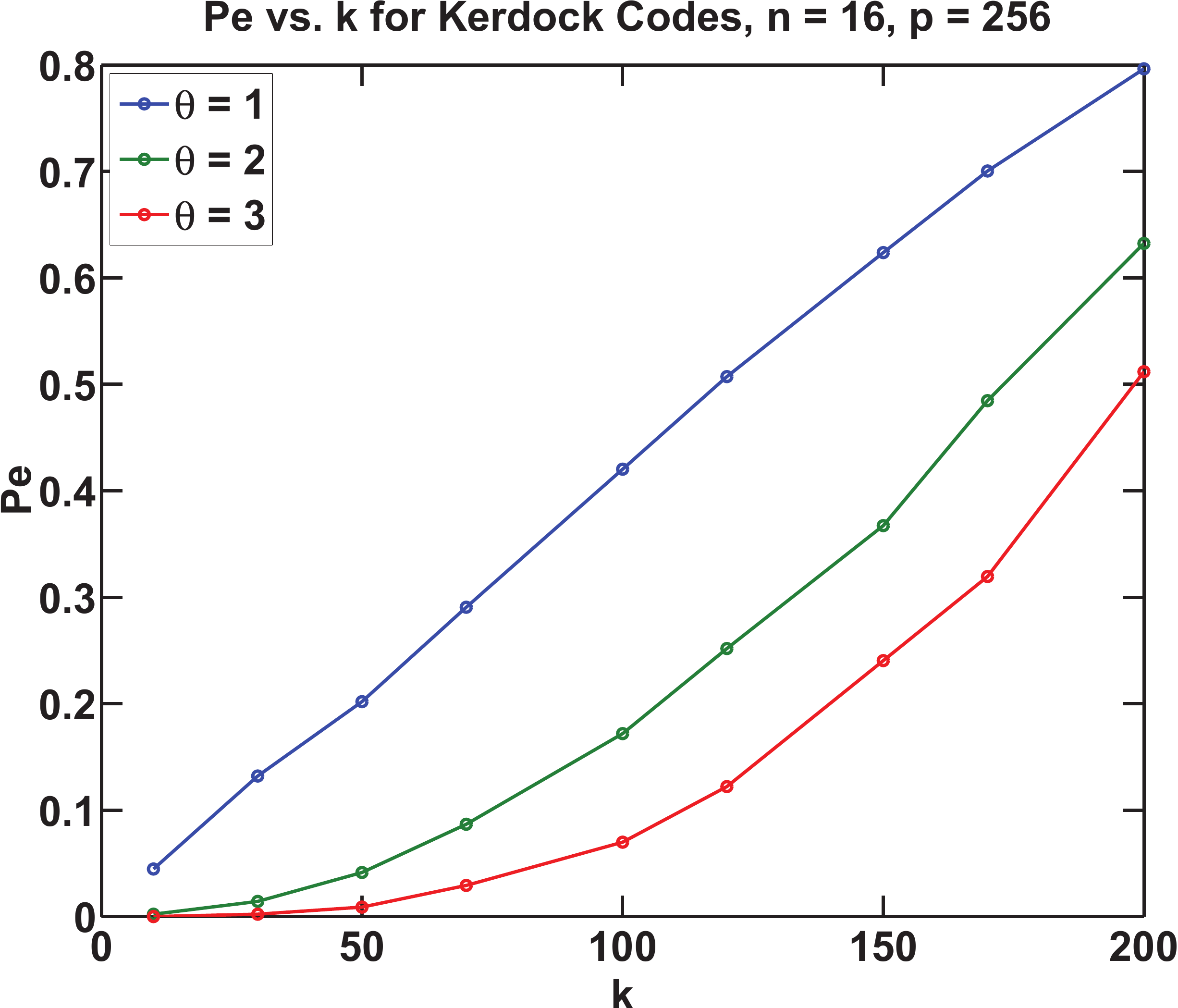}
\end{center}
\caption{$\text{P}_{\text{e}}$ versus $k$ for different choices of $\theta$ using Kerdock codes with dimension $16\times 256$. Each point was generated based on $5000$ independent trials. The simulation conditions were the same as those used in Fig.~\ref{fig:nongroupkerdockFDP}.\label{fig:nongroupkerdockPe}}
\end{figure}

\begin{figure}[h!]
\begin{center}
\includegraphics[width = 0.28\textwidth]{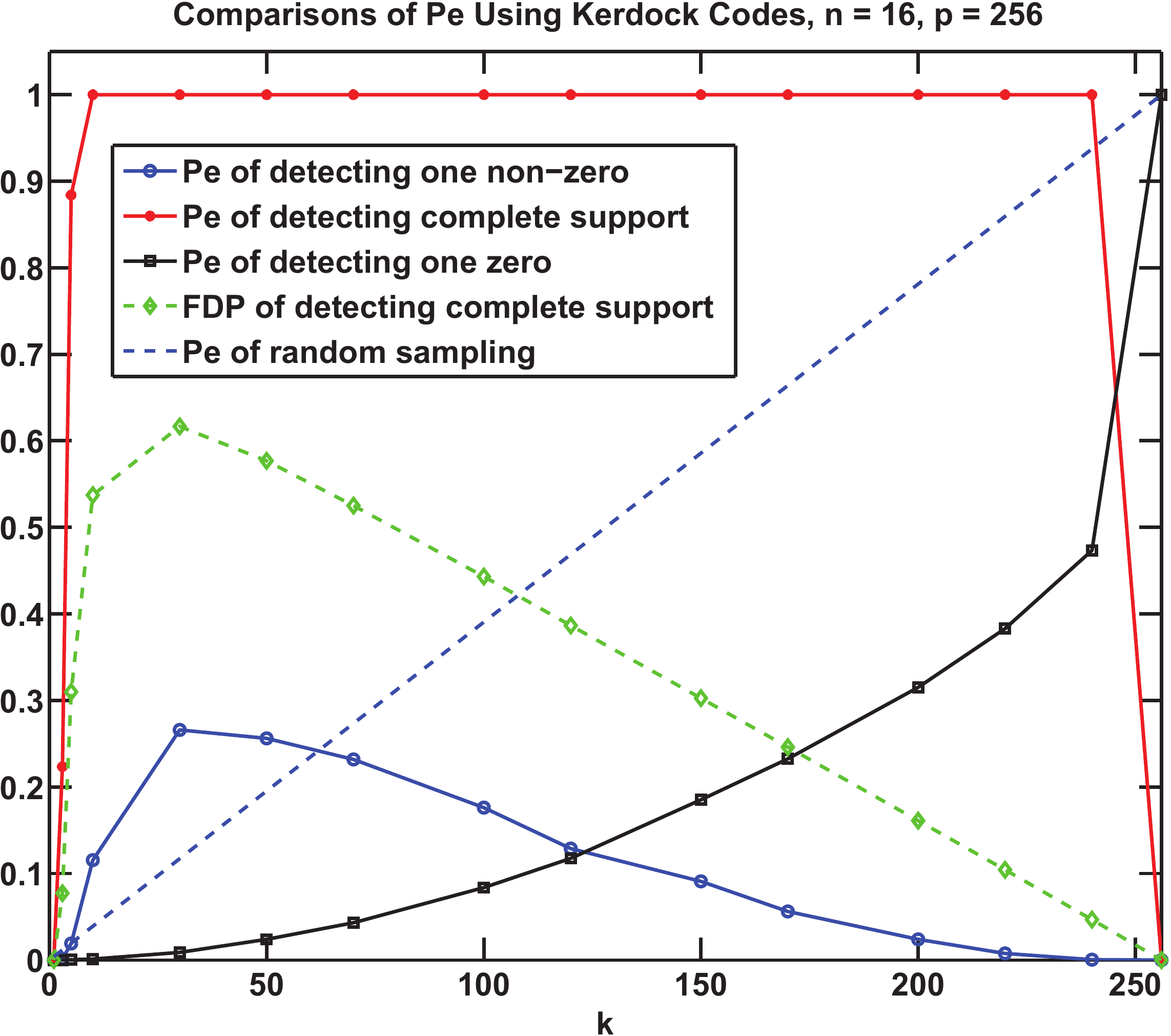}
\end{center}
\caption{A comparison of: the probability of error of detecting one zero $\text{P}_{\text{e}}$ 
 when $\theta = 1$, the probability of error of detecting one non-zero (using OST), and the probability of error and FDP of detecting the complete support. This example demonstrates that  
it is much easier to detect one zero than to recover the complete support, since in many scenarios, all we want is the location of ``one zero''.\label{fig:comparisons}}
\end{figure}

\begin{figure}[h!]
\begin{center}
\subfloat[Kerdock, group-thresholding]{
\includegraphics[width = 0.26\textwidth]{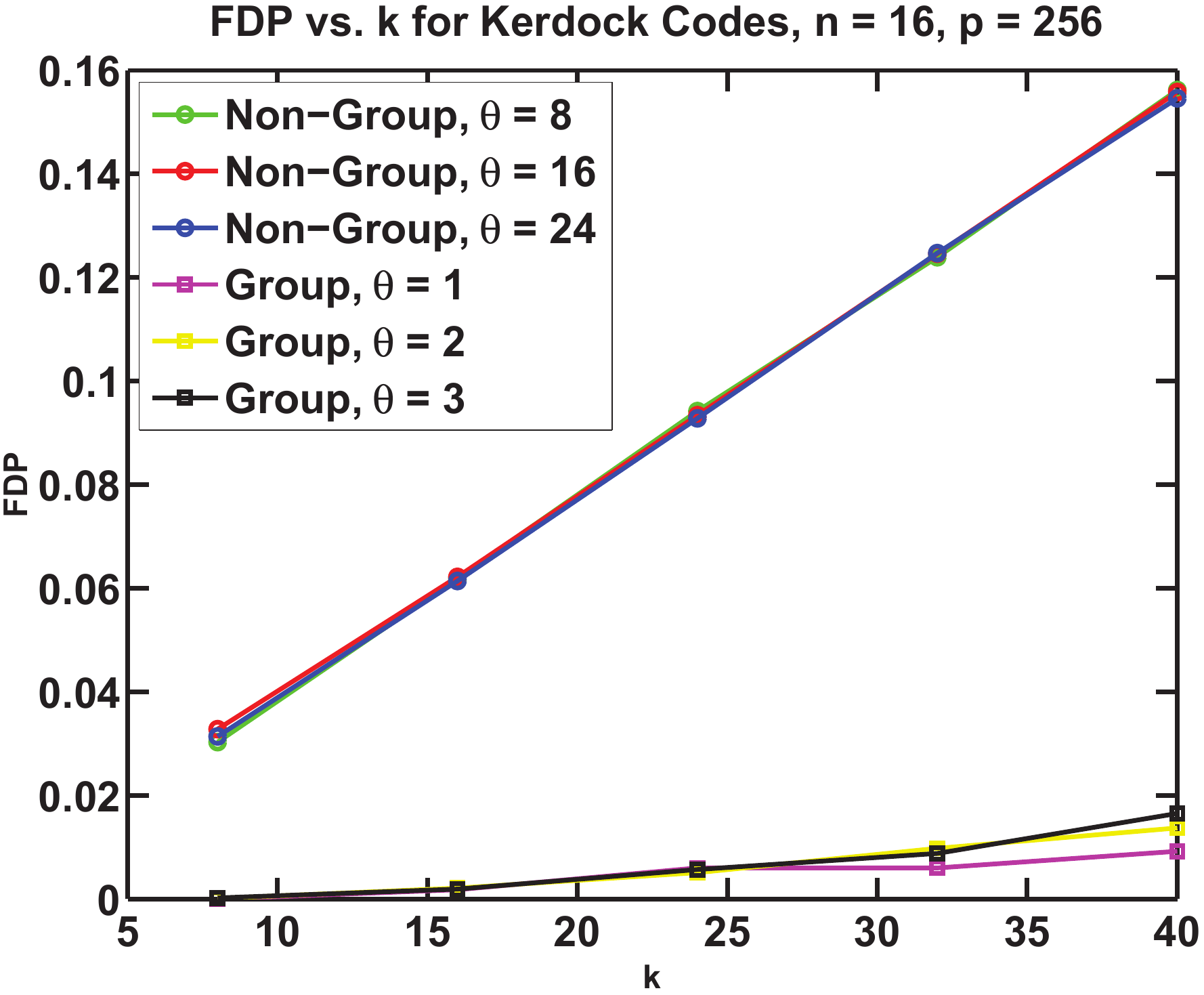}}
\quad
\subfloat[Bernoulli, group-thresholding]
{
\includegraphics[width = 0.26\textwidth]{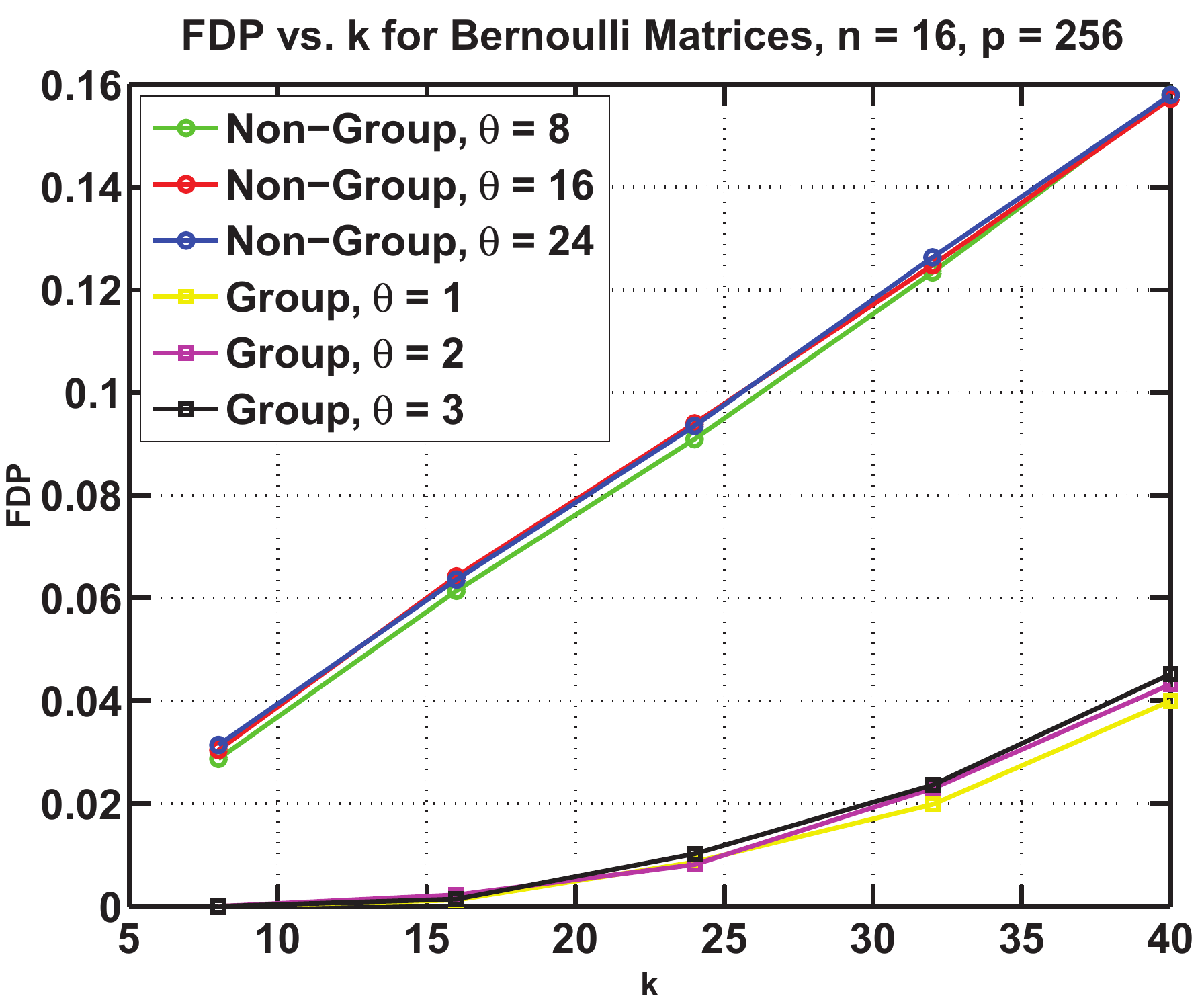}
}
\caption{A comparison of the performance of ZD-OST and ZD-GroTh for: (a) kerdock-preparata codes, and (b) the random bernoulli matrix. The comparisons are made for the same overall number of tones which is why the $\theta$ in the non-group data points are $8$ times the corresponding value of the Group data points.\label{fig:groupcomparisons}}
\end{center}
\end{figure}

This section experimentally demonstrates the efficacy of ZD-OST and ZD-GroTh at obtaining $\hat{\mathcal{E}}$ containing large subsets of the zero support.
We demonstrate the performance of ZD-OST and ZD-GroTh using both a random Bernoulli matrix 
and the $M\times M^2$ ($M=2^{m+1}$ with $m$ an odd integer) matrix of Kerdock-Preparata codes~\cite{kerdockz4} 
of dimension $16 \times 256$.  
The results are presented in terms of both FDP (Eq.~\ref{eq:fdp}) and $\text{P}_{\text{e}}$  as a function of the sparsity level of $\xb$ in the frequency domain $k=\norm{\mb{\beta}}_0 = \norm{F\xb}_0$. The input signal for tests of ZD-OST consisted of a superposition of $k$ tones from the DFT grid.
\beq
x_{\ell} =  \sum_{\omega \in \Omega, k = |\Omega|} a_{\omega} \e^{-2 \pi \iii \omega \ell}, \quad 
\Omega \subseteq \{ -(p/2 -1),\dots,p/2\}
\eeq
For the tests of ZD-GroTh, the random support consisted of randomly choosing $k$ groups of $r=8$ tones. Figures~\ref{fig:nongroupkerdockFDP}, \ref{fig:nongroupkerdockPe}, and~\ref{fig:comparisons} show results for ZD-OST and figures~\ref{fig:groupcomparisons}(a) and~\ref{fig:groupcomparisons}(b) show results for ZD-GroTh, including performance comparisons to ZD-OST.
Fig.~\ref{fig:nongroupkerdockFDP} shows the FDP performance of GroTh with respect to $k$ for several $\theta$. Also note, as $k$ becomes comparable to $p$, a large fraction of $\hat{\mathcal{E}}$ correspond to elements of the true zero-support. This would suggest that zero-detection would be amenable to use in an adaptive setting that would enable high-probability detection of zeros via remeasurement of the reduced set $\hat{\mathcal{E}}$. This is further evidenced by Fig.~\ref{fig:nongroupkerdockPe} which shows the $\text{P}_{\text{e}}$ performance for very low values of $\theta$.
Although the objectives differ considerably, it is illustrative to compare the $\text{P}_{\text{e}}$ (Fig.~\ref{fig:comparisons}) for different types of support recovery objectives via OST versus the $\text{P}_{\text{e}}$ of detecting a single zero when $\theta=1$. The $\text{P}_{\text{e}}$ for zero-detection remains considerably lower than its counterparts. While the $\text{P}_{\text{e}}$ is far worse in the regime of high $k$, in terms of applications like white-space detection, partial support recovery may not be as useful as partial zero-support recovery.
Figure~\ref{fig:groupcomparisons} illustrates that in the presence of group-structure, The FDP and $\text{P}_{\text{e}}$ performance of ZD-GroTh considerably outperforms ZD-OST. We also point out that the structured Kerdock-Preparata Codes also display superior performance to the random bernoulli matrix.
\vspace{-0.22cm}
\section{Conclusion}\label{sec:conclusion}
In this paper, motivated by the setting of white-space detection, we investigated using reduced-dimension measurements of a target signal $\xb$ to detect large subsets of its zero-support. Two algorithms, ZD-OST/ZD-GroTh, based on OST were presented to detect zeros in both the situation where group-structure is present and absent in $\xb$. Performance guarantees in terms of the probability of error and the FDP were proven in terms of the measurement matrix properties of worst and average coherence (group coherence). 
The performance of the algorithms was investigated empiricially using both measurement matrices based on random bernoulli and deterministic Kerdock-Preparata matrices. The numerical experiments demonstrated that  a high proportion of the detected zero-support sets ($\hat{\mathcal{E}}$) of even small cardinality ($\theta \ll p$) were elements of the true $\mathcal{E}$. We also note that even in regimes where the non-zero support is a large fraction of the signal dimension $k\sim 0.8p$, that a substantial fraction of $\hat{\mathcal{E}}$ contained elements of $\mathcal{E}$. 
We leave for future work extending our theory to cover the case of large $k$. Finally, we further point out that even in situations where detecting zeros is not the direct goal, efficient methods for finding zeros could still make considerable impact if they are incorporated into other recovery algorithms. For example, if methods for finding zeros are efficient and reliable, they could be used to improve the speed and cost of  computation by reducing the search space through quick determination of additional constraints in other recovery algorithms. 
\appendices
\section{Proof of Theorem 1}\label{sec:proof1}
\begin{proof}
Let $\tau \triangleq 2\sigma \sqrt{\log p}$. Define $\mathcal{G} = \{\max_{i=1}^p \|\ab^H \wb\| \leq \tau\}$. We can show that $\mathcal{G}$ occurs with probability at least $1-\sqrt{2/\pi}p^{-1}(\log p)^{-1/2}$.
To prove the first part of Theorem 1, note that when $\mathcal{G}$ occurs and StOC is satisfied,
\begin{equation}
\begin{split}
\min_{i\in\mathcal{E}}|\ab_i^H \yb| &= \min_{i\in\mathcal{E}}|\sum_{j} \ab_i^H \ab_j x_j + \ab_i^H \wb | \\
& \leq \min_{i\in\mathcal{E}}\left|\sum_{j} \ab_i^H \ab_j x_j\right| + \tau \\
& \leq \epsilon \|\xb\|_2 + \tau.
\end{split}
\label{bound_off_band}
\end{equation}
On the other hand, when $\mathcal{G}$ occurs and StOC: 
\begin{equation}
\begin{split} 
\min_{i\in\mathcal{I}}|\ab_i^H \yb| &=  \min_{i\in\mathcal{I}}\left|x_i + \sum_{j\neq i} \ab_i^H \ab_j x_j + \ab_i^H \wb \right|  \\
& \geq  \min_{i\in\mathcal{I}}|x_i| - \max_{i\in\mathcal{I}}|\sum_{j\neq i} \ab_i^H \ab_j x_j| - \tau \\
& \geq  |x_{\min}| - \epsilon \|\xb\|_2 - \tau.
\end{split}
\label{bound_on_support}
\end{equation}
Hence, when 
\ben
|x_{\min}| > 2\epsilon\|\xb\|_2 + 2\tau, \label{cond_11}
\een
$\min_{i\in \mathcal{E}}|\ab_i^H \yb| < \min_{i\in \mathcal{I}}\|\ab_i^H \yb\|$. This shows that under $\mathcal{G}$ and StOC, if (\ref{cond_11}) is satisfies, then for $\theta= 1$, $\hat{\mathcal{E}}\in \mathcal{E}$. 

In \cite{bajwaGabor2010}, it is shown that an $n\times p$ design matrix satisfies $(k, \epsilon, \delta)-$StOC for any $\epsilon \in [0, 1)$ with $\delta \leq 4p \exp\{-\frac{(\epsilon-\sqrt{k}\nu)^2}{16(2+a^{-1})^2\mu^2}\}$ for $a\geq 1$, $k\leq \min\{\epsilon^2\nu^{-2}, (1+a)^{-1}p\}$. Next we can choose proper parameters such that StOC. Substitute $\mu=\mu_0/\sqrt{\log p}$, we have that $\exp\{-\frac{(\epsilon-\sqrt{k}\nu)^2}{16(2+a^{-1})^2\mu^2}\} = p^{-\alpha}$, where $\alpha = (\epsilon-\sqrt{k}\nu)^2/(c\mu_0^2)$, where $c = 16(2+a^{-1})^2$. We want $\alpha > 1$ so that the bounds on probability of StOC is tight, which is satisfied when $k<(\epsilon-\sqrt{c}\mu_0)^2/\nu^2 < \epsilon^2/\nu^2$. Hence for these choice of parameters, we have that $\delta < 4p^{1-\alpha}$, $\alpha > 1$, when $k \leq \min\{(\epsilon-\sqrt{c}\mu_0)^2/\nu^2, (1+a)^{-1}p\}$, for a constant $a>1$. We want to choose the largest $\epsilon$ possible to make this bound tight, and from (\ref{cond_11}), for $\tau = 2\sigma\sqrt{\log p}$, the largest such $\epsilon_0 = (|x_{\min}|-2\tau)/(2\|\xb\|)=(\sqrt{\SNR_{\min}}-4\sqrt{\log p})/(2\sqrt{\SNR})$.

Combine the results above, we have that $\mathbb{P}\{\hat{\mathcal{E}}\in \mathcal{E}\}\geq \mathbb{P}\{\mathcal{G}\cap \mbox{StOC}\} \geq (1-\sqrt{2/\pi}p^{-1}(\log p)^{-1/2})(1-\delta) > 1-\sqrt{2/\pi}p^{-1}(\log p)^{-1/2}-4p^{1-\alpha}$. Thus the proof is finished by writing $P_e\leq 1-\mathbb{P}\{\hat{\mathcal{E}}\in \mathcal{E}\} < \sqrt{2/\pi}p^{-1} + 4p^{1-\alpha}.$

To prove the second part, notice that for $i \in \mathcal{I}$, similar to (\ref{bound_on_support}), we have that when $\mathcal{G}$ occurs and under StOC
\begin{equation}
\begin{split} 
|\ab_i^H \yb|
 \geq  |x_i| - \epsilon \|\xb\|_2 - \tau.
\end{split}
\label{bound_on_support}
\end{equation}
Hence if
\begin{equation}
|x_i| > 2\epsilon \|\xb\|_2 + 2\tau,  \label{cond_1}
\end{equation}
for $i \in \mathcal{I}$, we have that $|\ab_i^H \yb| > \max_{j\in \mathcal{E}} |\ab_j^H \yb|$, and hence $i\notin \mathcal{E}$. Suppose $m$ is the largest integer for which the following is true: $|x_{(m)}| > 2\epsilon \|\xb\|_2 + 2\tau$. Let $\ab_{(i)}$ correspond to the column of correspond to $x_{(i)}$.
Hence $|\ab_{(i)}^H \yb| > \max_{j\in \mathcal{E}} |\ab_j^H \yb|$ for $i = 1, \ldots, m$, $m\leq k$. Hence the number of components that are incorrectly detected can be at most $k-m$. Hence, we have
\ben
\FDP(\mathcal{E}) \leq (k-m)/\theta, \label{bound}
\een
when $\mathcal{G}$ occurs and StOC occurs. Finally, the theorem can be proved by noting that $|x_{(m)}| > 2\epsilon \|\xb\|_2 + 2\tau$ is equivalent to $|x_{(m)}| > 2\epsilon \|\xb\|_2/t$ and $|x_{(m)}| > 2\tau/(1-t)$ for $t\in (0, 1)$. As shown above, the probability that both $\mathcal{G}$ and StOC occurs is at least $1-\sqrt{2/\pi}p^{-1} - 4p^{1-\alpha}$. This finishes the proof.
\end{proof}
\section{Proof of Theorem 2}\label{sec:proof2}
Let  $\X_{\mathcal{K}}$ denote the $n\times rk$ sub-matrix of $\X$ that corresponds to the non-zero blocks, $\xb_{\mathcal{K}}$ denote the $rk\times 1$ sub-vector of $\xb$. Define $\tilde{\mathcal{K}} \triangleq \{i \in \mathcal{K}: \|\xb_i\|_2 \geq c_3 \mu^g \|\xb\|_2\sqrt{\log q}\}$.
 Then we have
\ben
\begin{split}
&\min_{i\in \tilde{\mathcal{K}}}\|\X_i ^H\yb\|_2 \\
&=\min_{i\in \tilde{\mathcal{K}}} \|\xb_i + (\X_i^H \X_{\mathcal{K}} \xb_{\mathcal{K}} - \xb_i) + \X_i^H \wb \|_2 \\
&\geq \min_{i\in \tilde{\mathcal{K}}} \|\xb_i\|_2 
-\max_{i\in \tilde{\mathcal{K}}} \|(\X_i^H \X_{\mathcal{K}}\xb_{\mathcal{K}} - \xb_i) \|_2-\max_{i\in \tilde{\mathcal{K}}}  \|\X_i \wb\|_2\\
&=\|\xb_{(L)}\|_2 - \|(\X_{\mathcal{K}}^H \X_{\mathcal{K}}-\I)\xb_{\mathcal{K}}\|_{2, \infty} + \max_{i\in \tilde{\mathcal{K}}}\|\X_i \wb\|
\end{split}
\een
We also have
\ben
\begin{split}
\max_{i\in \mathcal{K}^c} \|\X_i^H \yb\|_2 &\leq
\max_{i\in \mathcal{K}^c} \|\X_i^H \X_{\mathcal{K}} \xb_{\mathcal{K}}\|_2 + \max_{i\in \mathcal{K}^c}\|\X_i ^H \wb\|\\
& 
\end{split}
\een 
$\text{Hence, }\|\xb_{(L)}\|_2  > \|(\X_{\mathcal{K}}^H \X_{\mathcal{K}}-\I)\xb_{\mathcal{K}}\|_{2, \infty} + \max_{i\in \mathcal{K}^c} \|\X_i^H \X_{\mathcal{K}} \xb_{\mathcal{K}}\|_2 +\max_{i=1}^q\|\X_i \wb\|$
is a sufficient condition for $\min_{i\in \tilde{\mathcal{K}}}\|\X_i ^H\yb\|_2 > \max_{i\in \mathcal{K}^c} \|\X_i^H \yb\|_2$. 
Define $\tilde{\mathcal{G}} = \{\max_{i=1}^q \|\X_i^H \wb\|_2 < \tau\}$. Note that $\|\X_i^H \wb\|^2_2$ is a $\chi^2$ random variable with $r$ degrees of freedom. Using Chernoff bound, we have $\mathbb{P}\{\|\X_i^H \wb\|_2 > \tau\} \leq e^{-t \tau^2/\sigma^2} (1-2t)^{-r/2}$, for $t \in (0, 1/2)$. Choose $t = 1/4$, we have the lower bound: $e^{-\tau^2/(4\sigma^2)} 2^{r/2}$. From Sidak's Lemma, we have $\mathbb{P}\{\max_{i=1}^q\|\X_i^H \wb\|_2 < \tau\} \leq 1-qe^{-\tau^2/(4\sigma^2)} 2^{r/2}$. Let $\tau = (2\sigma\sqrt{2\log q + r/2\log 2} )$. This demonstrate that $\max_{i=1}^q\|\X_i^H \wb\|_2^2 < \tau$ with $\tau$ define above occurs with probability of at least $1-q^{-1}$. Combine this noise bound with [Proof of Theorem 1 in \cite{bajwaGroup}], we have that condition for correct detection occurs with probability of at least $(1-q^{-1})(1-e^2 q^{-1}) = 1+ (e^2+1)q^{-1} + o(q^{-1})$.
\bibliographystyle{IEEEbib}
\bibliography{refs}
\end{document}